\documentclass{aa}                   
\usepackage[dvips]{graphicx}         
\usepackage{graphics}
\usepackage{times}                   

\begin{document}

\thesaurus{02.05.1; 02.09.1; 02.14.1 06.1.2; 06.09.1}

\title{New type astrophysical solution to the solar neutrino problems and its predictions to the SNO}
\subtitle{}

\author{Attila Grandpierre}

\institute{
Konkoly Observatory of the Hungarian Academy of Sciences, H-1525 Budapest,
Hungary; grandp@konkoly.hu}

\date{Received \dots / Accepted \dots}

\maketitle

\begin{abstract}
The anomalously slow rotation of the solar core is just one from a remarkable lists of fundamental indications showing that the solar core is somehow coupled to the surface activity cycle. On the other hand, the atmospheric, LSND and solar neutrino problems are not consistent, therefore one or more of the neutrino experiments must be attributed - at least in part - to some phenomena other than neutrino oscillations, or a fourth neutrino is required. We use the luminosity constraint in a general case not considered yet, taking into account nuclear reactions occurring in small hot bubbles indicated to be present in the solar core (Grandpierre, 2000). The dynamic solar model fit the observed neutrino fluxes within one-and-half sigma even without oscillations. This fit is shown to be better than the present day MSW and VAC fits. An attracting perspective is obtained for a simultaneous solution of the neutrino problems and the solar core-related astrophysical problems. Predictions of the dynamic solar model are presented for the SNO measurements. 
 
\end{abstract}

\section{Introduction}
Two different approaches seem to be promising in exploring the roots of the solar neutrino problems. One is the traditional particle physics approach, suggesting that neutrino oscillations are responsible for the missing solar neutrinos. Unfortunately, the small mixing angle, large mixing angle and vacuum oscillation solutions all suffer from being poor fits to the observations. Eleven years ago Bahcall (1989) wrote in his book, that the MSW neutrino oscillation solution of the solar neutrino problem (SNP) is attractive since the mixing angles and mass differences can each vary by orders of magnitude, and so it does not need fine tuning as the vacuum solution does. But not so much later Paterno and Scalia (1994) noticed that the allowed region shrinkened to a point-like area, and so 
"hypotheses on non-conventional neutrino properties are strongly disfavoured, except for the matter neutrino oscillations, the latter surviving within very narrow limits". Now even this remained point-like area (and the surrounding small 95 $\%$ C.L. region, which extends only due to the theoretical and experimental errors) do not have a high allowance. The probability of these solutions decreased from 100 $\%$ to cca. 10$ \%$ for the widely regarded best fit of small mixing angle (SMA) solution (and similarly for the LMA and VAC solutions), when the SuperKamiokande (SK) rates, spectral and day-night effect data are also taken into account. 

One can evaluate the acceptability of the presented solutions by the probability $P(\chi^2)$ belonging to the calculated $\chi^2_{min}$. The rule of thumb telling that "the value of $ \chi^2$ for a "moderately" good fit is $ \chi^2 \approx d.o.f.$" (Press et al., 1992) may be useful. Bahcall, Krastev and Smirnov (1998) obtained for the SMA global fit $\chi^2_{min} = 26.5/17 d.o.f.$ which is acceptable at the probability $P(\chi^2)$ = 7 $ \%$ C.L.  Suzuki (1998) obtained $\chi^2_{min} = 50.2/31 d.o.f.$ for the case when oscillations are not allowed. The corresponding probability is $P(\chi^2) = 1.6 \%$. The best fit he obtained is for the vacuum oscillation (VAC), with $\chi^2_{min} = 38.7/31$ d.o.f., with $P(\chi^2) \approx 20 \% $. This latter case we recognize also as a solution worse than moderately good, since $\chi^2_{min} > d.o.f.$ At the same time, Bahcall, Krastev and Smirnov (1998) obtained a global fit (when the constraints from the rates, the spectrum shape and the Day-Night asymmetry are all included) to for the best fit vacuum oscillation solution $\chi_{min}^2=28.4$ for 18 d.o.f., which is acceptable only at 6$ \%$ C.L.
 
It may be disturbing that two different and contrasting confidences are introduced to characterise the acceptance of a fit. One is $P(\chi^2)$, the probability of a fit. The other one, $P( \Delta  \chi ^2)$, characterises how far the (allegedly normal) distribution of the experimental and theoretical errors extends. Since the larger errors leads to larger allowed regions, and the farther iso-probability contours belong to increasing $n$ characterising the distance as $n \sigma$, therefore the higher is this second probability, the worse is the fit. Therefore, the higher is the former probability $P(\chi^2)$, the better is the fit; while, on the contrary, the higher is the latter probability, $P( \Delta  \chi ^2)$, the worse is the fit. 

In Gonzalez-Garcia and Pena-Garay (2000) Table 3, the larger C.L. belongs to a higher $\chi^2/d.o.f$ and so it shows a poorer fit. They found for the SMA a C.L. $83 \%$, which tells us that it is somewhere outside the 1.5 $\sigma$ region around the best fit which has a low (cca. 15 $\%$) probability itself. Langacker (1999) noted that the LMA solution and the no oscillation hypothesis has the same rate of goodness, i.e. both of them is disfavored at the 95-99 $ \%$ CL level, while the LMA solution is also a very poor fit, although it is allowed at 95$\%$ CL. In other terms, this means that SMA is refuted with more than 1.5 $\sigma$. 

Recently Maris and Petcov (2000) found that "the conservative SMA around the point $ \Delta m^2 = 4 \times 10^{-6}$, $sin^2 2 \Theta = 0.0085$ is ruled out at 1.5 sigma" by the degree of the day-night effect observed by the SK. 
Now Haxton (2000) noted that "One puzzling aspect of atmospheric, solar, and LSND neutrino results is that they require three independent $ \delta m^2$s. That is, they do not respect the relation $ \delta m^2_{21} +  \delta m^2_{31} + \delta m^2_{32} = 0 $, thus either one or more of the neutrino experiments must be attributed to some phenomena other than neutrino oscillations, or a fourth neutrino is required." 

We learned that the neutrino oscillation solutions of the SNP actually are poor fits. Moreover, we recognise that even if they would be moderately good fits, it would be still an urgent need to find new physics outside from the neutrino oscillations to explain the results of the neutrino experiments. In this Letter we suggest to look for the new physics in a conservative region, in the field of astrophysics. The reason is the (not yet recognised) significance of the facts corresponding to the coupling of the solar core to the surface activity phenomena (one can find the presentation of the astrophysical core-related problems in Grandpierre, 1996, 1999, 2000). It is known, that the solar core rotates so slowly that its rotation rate is the surface rate $\pm 30 \%$ (Elsworth et al., 1995). At the same time, the so-called 'best solar model' of Pinsonneault et al. (1989) predicted a rotation rate of the core which is 4-15 times the surface rate. The discrepancy may be surmounted only when allowing a coupling between the solar core and the surface active regions. But if the core may participate in the surface activity cycle, the energy production has to be touched, and so the neutrino production may be influenced as well. In this way we found a way how the new physics arises in the field of astrophysics. 

The objections raised against a possible astrophysical solution to the solar 
neutrino problems (see e.g. Bludman, Hata and Langacker 1994) are valid only for non-standard models without new physics (without changing the spectral shape of the individual neutrino fluxes). But it is known that the solution of the SNP needs new physics (Bludman, Hata and Langacker, 1994). The often used term "model-independence" refers to general models in which the individual pp, $Be^7$, CNO and $B^8$ neutrino fluxes are allowed to vary as free parameters. But if new physics is present, another kind of nuclear reactions may also contribute to the energy and neutrino production of the Sun. In this paper we attempt to show how the presence of a high-temperature energy source (Grandpierre, 1996, 2000) could contribute to the neutrino detector data.

It is a general view that the principal neutrino sources are pp, pep, $^7Be$, $^8B$, $^{13}N$, and $^{15}O$. As Bahcall and Krastev (1996) remarked, this picture has become so accepted that it is sometimes referred to as "model-independent". Nevertheless, if other neutrino sources do exist, then these previously thought "model-independent" models may all fail at a certain rate. Therefore, the related luminosity constraint also may prove insufficient and over-constrained when compared to the actual Sun, since the hot bubbles may produce a smaller or larger part of the solar luminosity. 
The luminosity constraint expresses the fact that the energy productions is related to the neutrino production since both is produced by the nuclear reactions:
\begin{eqnarray}
L_{\odot}/4\pi R^2 =  \sum_{\alpha} (Q/2 - <E>_{\alpha} ) \phi(\alpha),
\end{eqnarray}
where R=1 A. U. (1.469$\times 10^{13}$ cm), $<E>_{\alpha}$ is the average neutrino energy and $\phi(\alpha)$ is the $\alpha$th flux ($\alpha$=pp, pep, $^7Be,^8B,$...), and $Q$ is the energy released in the fusion reaction $4p+2e \rightarrow \alpha + 2\nu$.
The most general luminosity constraint we found in the literature is presented in Minakata and Nunokawa, 1999 in the following form:
\begin{eqnarray}
L_{\odot}/4\pi R^2 & = 13.1 \phi(pp) + 11.92 \phi(pep) + 12.5 \phi(^7{Be})+ \nonumber\\
 &  6.66 \phi(^8B) + 3.46 \phi(^{13}N)+21.57 \phi(^{15}O)\nonumber\\ 
 & + 2.36 \phi(^{17}F) + 10.17 \phi(hep).
\end{eqnarray}
We found it more suitable to convert these neutrino fluxes to fractional fluxes $\Phi$, normalised to the BP98 SSM fluxes:
\begin{eqnarray}
1 & = 0.9119 \Phi(pp) + 0.001966 \Phi (pep) + 0.0758 \Phi(Be) \nonumber\\
 & + 5.193 \times 10^{-5} \Phi(B)+ 3 \times 10^{-3} \Phi(CNO). 
\end{eqnarray}
Neglecting the small terms, and including the pep into the pp, the CNO into the Be term,
\begin{eqnarray}
0.99= 0.914 \Phi_1 + 0.076 \Phi_7, 
\end{eqnarray}
where $ \Phi_1 $ refers to the pp+pep fluxes, and  $\Phi_7$ to the Be+CNO fluxes (where we modified the left side with the condition that for  $\Phi_1$=$ \Phi_7$=1 the equation must be valid). 

We used the fractional neutrino fluxes following Minakata and Nunokawa (1998). The chlorine equation is:
\begin{eqnarray}
2.56 = 1.8 \Phi_7 + 5.9 \Phi_8.
\end{eqnarray}
The gallium-equation is: 
\begin{eqnarray}
72.4 = 69.6 \Phi_1 + 46.9 \Phi_7 + 12.4 \Phi_8. 
\end{eqnarray}

\begin{figure}[t]
\includegraphics[angle=0,width=7.5cm]{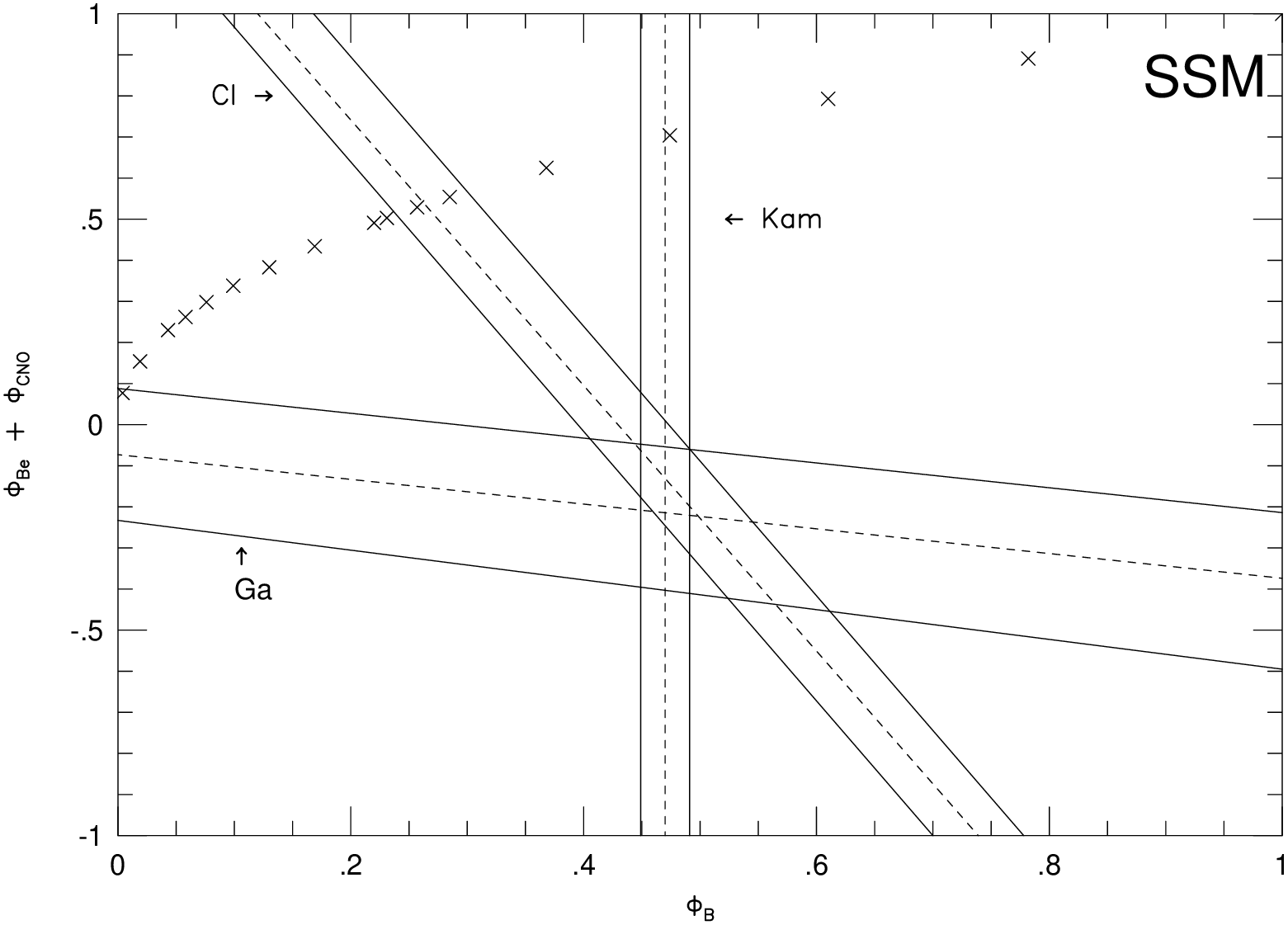}
\caption[ ]{The $^8B$ and $^7Be$+CNO neutrino fluxes, consistent with the luminosity constraint and experimental results for standard neutrinos. The SSM flux region is around the (1,1) point. The dashed (solid) lines correspond to the central ($\pm 1 \sigma$) experimental values for Cl, Ga and $\nu -e$ scattering experiments. The crosses indicate the behaviour of non standard solar models with low central temperature. }
\label{F1}
\end{figure}

\begin{figure}
\includegraphics[angle=0,width=7.5cm]{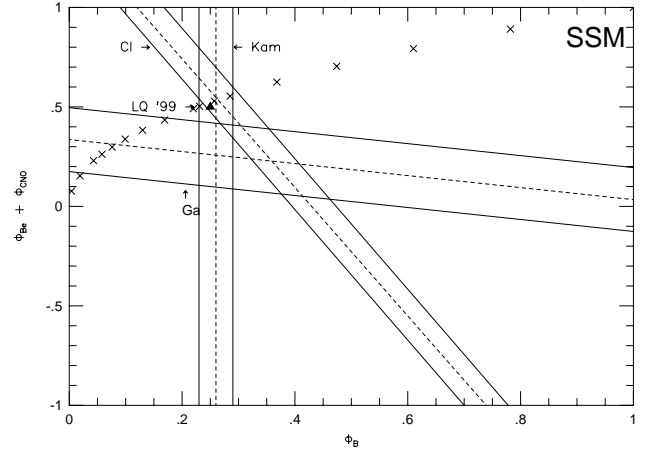}
\caption[ ]{The same as in Fig. 1, except that the luminosity constraint is now written for the quiet solar core of the dynamical solar model. The hot bubbles are allowed to produce mu and tau neutrinos, therefore the Kam line shifts leftwards. The triangle LQ'99 refers to the calculations of Lavagno and Quarati, 1999. The figure shows that LQ'99 does not fit exactly to the low-temperature parabola.}
\label{F2}
\end{figure}

It is easy to derive from these equations the lines of the Fig. 1 of Fiorentini and Ricci (1998), shown here as Fig. 1. It is confirmed that the non standard cool Sun models do not give an acceptable fit, as they are farther than $3 \sigma$ from the intersection of the acceptance zones.

Now we are prepared to consider the case when hot bubbles are present. If the energy production of the hot bubbles is not negligible when compared to the total solar luminosity, we should involve it into the luminosity constraint. Regarding the energy budget of the hot bubbles, it depends on their temperature, which is indicated to be in the range of $10^8$ to $10^{11}$ K (Grandpierre, 1996, 2000). 
One can think that the bubbles may produce energy through the hot CNO cycle, triple alpha cycle (this cycle does not produce neutrinos) and other nova-type nuclear reactions (Audouze, Truran, Zimmerman, 1973). From these reactions the one which may produce the largest number of events in the neutrino detectors may be the hot CNO cycle. Writing the luminosity constraint for the hot bubble separately, for the case when the CNO cycle gives ten percent of the total solar luminosity as an upper limit, from (1) we obtain:
\begin{eqnarray}
0.1 L_{\odot}= 12.525 \Phi^b(CNO).
\end{eqnarray}
From this constraint, we can derive an upper limit for the bubble CNO neutrino flux,
\begin{eqnarray}
\Phi^b(CNO) < 6 \times 10^9 cm^{-2} s^{-1}, 
\end{eqnarray}
which is close to the value of the SSM beryllium-neutrino flux. This value is compatible with the present-day global constraints $0.0 < \Phi(^7 Be) < 6.35$ (Bahcall and Krastev, 1996). 

Now we can calculate how Fig. 1 is modified when the hot bubbles are producing 22$ \%$ of the total solar luminosity. In this case the luminosity constraint for the quiet solar core without the bubbles should be formulated as:
\begin{eqnarray}
0.77= 0.914 \Phi_1 + 0.076 \Phi_7. 
\end{eqnarray}
Using this constraint in the Ga-equation (6), we observe that the effect of the hot bubble energy generation, or, more precisely, the constraint that the quiet solar core should produce a less than total solar luminosity, is to shift the Ga-zone upwards. This effect is helpful in obtaining better and physical (i.e. $\Phi(Be) > 0$) fluxes, and so to resolve one of the solar neutrino problems, the problem of beryllium flux.

Now one may consider what happens with the Kamiokande zone if hot bubbles are present. Since the bubbles may reach very high temperatures, they may produce mu and tau neutrinos, which can be observed by the Kamiokande and not by the other detectors. Therefore, since Fig. 2 describes the quiet solar core only, the Kam-zone may be shifted to the left. Actually, the rate of the necessary shift is determined by the intersection of the Cl and Ga zones. We observe from Fig. 2 that the shift occurs around $\Phi_8$=0.26, which means that the bubbles has to produce a $\nu_{\mu, \tau}$ neutral current with a contribution to the SK 
\begin{eqnarray}
\Phi^b_{\mu, \tau} = 0.21. 
\end{eqnarray}
We note here, that since the hot bubbles may contribute also to the high-energy excess observed at the SK, therefore the amount of $\Phi^b_{\mu, \tau}$ should not be so large. The Cl-zone of Fig. 1 will not be modified when going to Fig. 2. 

\section{Discussion and Conclusions}
It is interesting that the dynamic solar model (DSM) obtained with the inclusion of the hot bubbles into the standard solar model (SSM) modifies the Ga, Cl, Kam and non-standard zones in a way the create shifting an overlap region in a physical range ($\Phi_7 > 0$. Remarkably, the cool Sun model also overlaps with this overlap region. For a bubble-luminosity around 0.22 (of the total solar luminosity), they all fit with cca. 1.2 $\sigma$. Fig. 2 shows that the three kinds of neutrino detectors actually do not contradict to the standard neutrino picture. The situation is that the combined neutrino results actually indicate a lower than standard central temperature of the Sun with a remarkable confidence. Fig. 2 gives a better fit (cca. 1.2 $\sigma$) than that of the presently favoured SMA solution (higher than 1.5$\sigma$). The best fit may be reached when both the bubbles and the oscillations are taken into account. 

One can observe from Fig. 2 that the new astrophysics may offer powerful perspectives to influence the solution of the SNP. It seems to be not true, that we know the basic physics of the Sun enough. Even if one can consider the presence of the hot bubbles in the solar core as yet not established, the frequent statements that astrophysical solutions are ruled out, prove to be unfounded in the here presented more general basis. 

In Fig. 2 the non standard low-temperature solutions are also presented. Although in Fig. 1 they are found far from the 1 $\sigma$ zones, in the case of Fig. 2 they are around 1.2 $\sigma$ from the overlap region when $T_c = 0.942$. This value of central temperature is quite consistent with the 0.78 solar luminosity produced by the quiet solar core.

We do not attempt to suggest that the hot bubbles permanently produce a significant part of the solar luminosity. Our attempt is more narrowly confined: it is to point out that the bubbles in principle may create a situation in which the astrophysical solution may be {\it alone} enough to solve the problem of the missing solar neutrinos. It is quite plausible, that the electron neutrinos do oscillate, and this phenomenon is responsible for a large part of the SNP and other apparent "neutrino anomalies". Nevertheless, the point is that if we ignore the role played by the dynamics of the solar core in the production of solar neutrinos, we may found ourselves in the uncomfortable situation of poor fits of the MS parameters and inconsistent consequences of atmospheric, LSND and solar neutrino problems. 

To resolve the astrophysical problems of the solar core it is not necessary that the hot bubbles produce a significant part of the solar luminosity. Actually, they may produce only a negligible part of the solar luminosity and they could prove still be able to trigger and influence the surface solar activity. Gorbatzky (1964) already calculated that hot bubbles arising from point explosions with an initial surplus energy around $10^{35}$ ergs may be able to reach the stellar surfaces from 0.1 stellar radius. He ignored completely the energy production of these hot bubbles with temperatures $T > 10^8 $ K. But even a slight influence of the solar core dynamics might be able to improve significantly the fit and consistency of the neutrino problems.

The result red from Fig. 2 has predictive value for the future neutrino detectors. It is easy to estimate the consequences of this dynamic solar model (for standard neutrinos) regarding the SNO observations. {\it Our picture modifies the conclusion that the [NC]/[CC] is larger than unity} ([i]=observed rate/standard solar model rate, see Bahcall, Krastev, Smirnov, 2000) {\it will definitely indicate the presence of neutrino observations}. Since at high temperatures like $ 10^{10} - 10^{11} K$ the hot bubbles may produce mu and tau neutrinos independently of the presence of neutrino oscillations, they may contribute to [NC] increasing it. For example, if we take $T=0.942$, than we will have $R^{qc}_{SK}=0.23$, and so [NC]/[CC]=2.04. If we allow a value of T closer to the SSM value 1, the ratio $([NC]/[CC])_{DSM}$ decreases towards unity. This conjecture of the DSM should be taken into account at the interpretations of the SNO and SK observations. 

Regarding the helioseismic context, we note that sound speed is not sensitive to the nuclear reactions. Bahcall and Ulrich (1988) and Basu, Pinsonneault and Bahcall (2000) remarked that even when switching out the $He^3$ + $He^4$ reaction (producing 14 $\%$ of the total solar luminosity), the sound speed differ only by 0.1 $\%$ from the sound speed obtained from the standard solar model. On the other hand, the calculations of the dynamic solar model show that when bubbles and neutrino oscillations are both present the SMA and LMA regions (of the purely MSW solutions) shift significantly even for such a small change as $T=0.995$. Moreover, the many-body effects of the particles in the nuclear reactions led to solar models which may compensate the cool solar models towards an SSM central temperature (Lavagno, Quarati 1999).


\end{document}